\documentclass[3p,twocolumn]{elsarticle}

\usepackage{amsmath,amsfonts,amssymb}
\usepackage{graphics}
\usepackage{subfigure}
\usepackage{color}

\newcommand{\href}[2]{#2 ({\tt #1})}
\newcommand{\url}[1]{{\tt #1}}

\newcommand{\newstuff}[1]{#1}

\newenvironment{mat}{\left[ \begin{array}{ccccccccccccccc}}{\end{array}\right]}
\newenvironment{rmat}{\left[ \begin{array}{rrrrrrrrrrrrr}}{\end{array}\right]}
\def\bcm{\begin{mat}}   
\def\ecm{\end{mat}}
\def\brm{\begin{rmat}}  
\def\erm{\end{rmat}}

\newenvironment{pvect}{\left( \begin{array}{c}}{\end{array}\right)}
\def\bpvect{\begin{pvect}}
\def\epvect{\end{pvect}}

\newenvironment{pwdef}{\left\{ \begin{array}{ll}}{\end{array}\right.}
\newcommand\bpwdef{\begin{pwdef}}
\newcommand\epwdef{\end{pwdef}}
\newcommand\when{&\m{if}}

\newcommand{\m}[1]{~~~\mbox{#1}~\,}

\def\bsplit{\begin{split}}
\def\esplit{\end{split}}

\newcommand{\Fig}[1]{Figure~\ref{fig:#1}}
\newcommand{\Sec}[1]{Section~\ref{sec:#1}}

\newcommand{\ignore}[1]{}
\newcommand{\Comment}[1]{}
\newcommand{\eqn}[1]{(\ref{#1})}
\newcommand{\eq}{\begin{equation}}
\newcommand{\en}{\end{equation}}
\newcommand{\eqm}{\begin{eqnarray}}
\newcommand{\enm}{\end{eqnarray}}
\newcommand{\eqmno}{\begin{eqnarray*}}
\newcommand{\enmno}{\end{eqnarray*}}
\newcommand{\eqml}[1]{\eql{#1}\begin{array}{rcl}}
\newcommand{\enml}{\end{array}\en}
\newcommand{\eql}{\begin{equation}\label}
\newcommand{\eqsub}[1]{\begin{subequations}\label{#1}\eqm }
\newcommand{\ensub}{\enm\end{subequations}}
\newcommand{\half}{{\frac{1}{2}}}

\def\bc{\begin{center}}
\def\ec{\end{center}}
\def\bi{\begin{itemize}}
\def\ei{\end{itemize}}
\def\be{\begin{enumerate}}
\def\ee{\end{enumerate}}

\def\reals{{{\rm l} \kern -.15em {\rm R} }}

\def\qquad{\quad\quad}

\ignore{

\newcommand{\ignore}[1]{}

\newenvironment{mat}{\left[ \begin{array}{ccccccccccccc}}{\end{array}\right]}
\newenvironment{rmat}{\left[ \begin{array}{rrrrrrrrrrrrr}}{\end{array}\right]}
\newenvironment{lmat}{\left[ \begin{array}{lllllllllllll}}{\end{array}\right]}
\newcommand\bcm{\begin{mat}}
\newcommand\ecm{\end{mat}}
\newcommand\brm{\begin{rmat}}
\newcommand\erm{\end{rmat}}
\newcommand\blm{\begin{lmat}}
\newcommand\elm{\end{lmat}}

\newcommand\bc{\begin{center}}
\newcommand\ec{\end{center}}
\newcommand\bi{\begin{itemize}}
\newcommand\ei{\end{itemize}}
\newcommand\be{\begin{enumerate}}
\newcommand\ee{\end{enumerate}}

\newcommand{\m}[1]{~~~\mbox{#1}~\,}
\newcommand\bsplit{\begin{split}}
\newcommand\esplit{\end{split}}
\newcommand{\eqn}[1]{(\ref{#1})}
\newcommand{\eq}{\begin{equation}}
\newcommand{\en}{\end{equation}}
\newcommand{\eqm}{\begin{eqnarray}}
\newcommand{\enm}{\end{eqnarray}}
\newcommand{\eqmno}{\begin{eqnarray*}}
\newcommand{\enmno}{\end{eqnarray*}}
\newcommand{\eqml}[1]{\eql{#1}\begin{array}{rcl}}
\newcommand{\enml}{\end{array}\en}
\newcommand{\eql}{\begin{equation}\label}
\newcommand{\eqsub}[1]{\begin{subequations}\label{#1}\eqm }
\newcommand{\ensub}{\enm\end{subequations}}

\newcommand{\half}{{\frac{1}{2}}}

\newcommand\reals{{{\rm l} \kern -.15em {\rm R} }}

\newcounter{equationgroup}
\newcommand{\ico}[1]{#1 \kern -1ex \raisebox{1.1ex}{$\circ$}}
\newcommand{\icos}[2]{#1 \kern -1.1ex \raisebox{1.1ex}[.5em]{$\circ$}^{#2}}

\newcommand{\Proj}{{{\rm l} \kern -.15em {\rm P} }}

}

\begin{document}
\title{ The GeoClaw software for depth-averaged flows
with adaptive refinement}

\author{
Marsha J.\ Berger,
David L.\ George,
Randall J. LeVeque, and
Kyle T. Mandli%
}

\address{Courant Institute of Mathematical Sciences, NYU, \\
Cascades Volcano Observatory, U.S. Geological Survey,
1300 SE Cardinal Ct. \#100, Vancouver, WA 98683\\
Department of Applied Mathematics,
University of Washington, Box 352420, Seattle, WA 98195-2420}

\date{July 31, 2010}

\begin{abstract}
Many geophysical flow or wave propagation problems can be modeled with
two-dimensional depth-averaged equations, of which the shallow water
equations are the simplest example.
We describe the GeoClaw software that has been designed to
solve problems of this nature, consisting of open source Fortran programs
together with Python tools for the user interface and flow visualization.
This software uses
high-resolution shock-capturing finite volume methods on logically
rectangular grids, including latitude--longitude grids on the sphere.
Dry states are handled automatically to model inundation.
The code incorporates adaptive mesh refinement to allow the
efficient solution of large-scale geophysical problems. Examples are given
illustrating its use for modeling tsunamis and dam-break flooding problems. Documentation and download information is available at
{\tt www.clawpack.org/geoclaw}.
\end{abstract}

\maketitle

\section{Introduction}\label{sec:intro}
Many geophysical flow or wave propagation problems take place over very
large spatial domains, for which
detailed three-dimensional modeling of the fluid
dynamics is not an efficient option.  Fortunately,
two-dimensional depth-averaged equations such as the shallow water
equations often provide models that are sufficiently accurate for many
applications.
Even with two-dimensional models, however, it is often necessary to
use adaptive mesh refinement (AMR) techniques in order to
concentrate grid cells in regions of interest, and to follow such regions as
the flow evolves.  This is often the only efficient way to
obtain results that have sufficient spatial resolution where needed
without undue refinement elsewhere, such as regions the flow or
wave has not yet reached or points distant from the study area.

We will briefly describe and illustrate the use of GeoClaw, an open source
research code that uses high-resolution finite volume methods together with
adaptive mesh refinement to tackle geophysical flow problems. In particular,
this code has recently been used together with the shallow water equations to
model tsunamis and dam-break floods. In \Sec{apps} we give a
brief illustration of each. For other geophysical flow problems it may be
necessary to replace the shallow water equations by a different set of
depth-averaged equations. For example, in modeling landslides, debris flows, or
lahars, it is necessary to incorporate terms modeling internal stress or pore
pressure (e.g. \cite{DenlingerIverson2004, PelantiBouchutEtAl2008a}). The
software is written in a manner that allows such extensions.

GeoClaw is based on the Clawpack software and is incorporated as
a part of the general Clawpack distribution \cite{claw.org.url}.
Clawpack (Conservation Laws Package) is an open source software
package that has been under development since 1994 and is widely used for
both teaching and research purposes.  It is designed to solve
hyperbolic systems of partial differential equations (PDEs) in
one, two, and three space dimensions.
This class of PDEs generally models wave propagation
or fluid transport, and a wide variety of physical problems give rise to
mathematical models of hyperbolic form, including for example compressible
gas dynamics, linear and nonlinear acoustics, and elastic wave propagation.
The theory of nonlinear hyperbolic systems and
a variety of applications are described in \cite{rjl:fvmhp}, which
also describes in detail the high-resolution finite volume methods
implemented in Clawpack.  Nearly all of the examples given in this text are
available as working examples via the Clawpack website.

Clawpack is written in a formulation that allows the user to specify the
system of equations being solved by providing a ``Riemann solver'' as
described in \Sec{num}. The software incorporates a general form of AMR as
reviewed briefly in \Sec{amr}, in a manner that is easy to apply to many
hyperbolic problems.  However, there are several difficulties that
arise when solving depth-averaged equations over realistic topography
or bathymetry that required some substantial modifications to the
general approach taken in Clawpack. The GeoClaw variant of the code
provides an implementation specific to such problems.

In particular, this code addresses the following issues:
\begin{itemize}
\item The flow takes place over  topography or bathymetry that may be
specified via multiple data sets covering overlapping regions at different
resolutions.  (Henceforth we will generally use the term topography to refer also
to bathymetry.)
\item Some problems can be tackled on
purely Cartesian grids, but many applications require using
longitude--latitude grids on the earth's surface.
\item The flow is of bounded extent; the depth goes to zero at the margins
and the ``wet-dry interface'' is a moving boundary that must be captured as
part of the flow.  This is handled by allowing the fluid depth to be zero in some grid
cells (``dry cells'').  Cells can change dynamically between wet and dry to model
evolving flows or inundation, and AMR can be used to provide sufficient
resolution of the shoreline or margin.
\item There often exist nontrivial steady states (such as an ocean at
rest) that should be maintained exactly.
Often the desired flow or wave propagation is a small perturbation of this
steady state, as in tsunamis.
\newstuff{For finite volume methods that conserve mass by using the
depth as a primary variable, this}
requires the use of a ``well-balanced'' numerical method as discussed
in \Sec{num}.
\end{itemize}
These issues and the algorithms in GeoClaw are discussed in more detail
elsewhere \cite{George2006, George2008, GeorgeLeVeque2006,
LeVequeGeorge2008,mjb-dg-rjl:actanum2011} and here we give only a brief summary of some key
aspects of the numerical algorithms (in \Sec{num}) and the AMR procedure (in
\Sec{amr}).

The computational core of GeoClaw is written in Fortran, but a user
interface written in Python is provided to simplify the setup of a single
run, or of a series of runs as is often required for parameter studies,
sensitivity studies, or probabilistic assessments of hazards.  Python
and Matlab
plotting tools are also provided for viewing the results in various forms,
either on the dynamically changing set of adaptive grids or on a set of
fixed grids, or in other forms such as gauge plots of depth vs.~time at
fixed spatial locations.  Some of these software tools are described briefly
in \Sec{software}, and more details can be found in the on-line documentation
\cite{geoclaw-doc}.

\section{Depth-averaged mathematical models}\label{sec:eqns}

The simplest depth-averaged set of
fluid equations in two lateral space dimensions are the
shallow water equations
\begin{equation}\label{2dSWE}
\begin{split}
  h_t &+ (hu)_x + (hv)_y = 0,\\
  (hu)_t &+ (hu^2 + \half gh^2)_x + (huv)_y = -ghB_x - Du,\\
  (hv)_t &+  (huv)_x+ (hv^2 + \half gh^2)_y = -ghB_y - Dv,
\end{split}
\end{equation}
where $u(x,y,t)$ and $v(x,y,t)$ are the depth-averaged velocities in the two
horizontal directions, $B(x,y,t)$ is the topography or bathymetry,
and $D=D(h,u,v)$ is the drag coefficient.
Coriolis terms can also be added to the momentum equations.
The equations \eqn{2dSWE} have the form
\eql{2deqn}
q_t + f_1(q)_x + f_2(q)_y = \psi(q,x,y)
\end{equation}
where $q = (h,~hu,~hv)$ is the vector consisting of the depth and momentum of the fluid.
In the absence of bathymetry ($B\equiv $ constant, so $B_x=B_y=0$) and drag ($D\equiv
0$), the source terms would be zero ($\psi\equiv 0$) and
these equations would express the conservation of mass and horizontal momentum.
We use conservative finite volume methods that in general conserve mass to
machine precision (since there is no source term in the mass equation) and
would also conserve momentum in the absence of source terms.  This is true
even when AMR is applied, with the exception of cells that intersect the
coastline, as discussed further in \Sec{amr}.

Note that for an ocean at rest, in which $h(x,t)+B(x,y) \equiv 0$ (sea level) in
all wet cells,
the topography source terms exactly cancel the derivatives of the hydrostatic pressure
$\half gh^2$.  Maintaining this balance numerically is critical and is discussed in
\Sec{num}.  The drag term could have many forms; for the experiments reported here we
use
\eql{drag}
D = \frac{gM^2\sqrt{(u^2+v^2)}}{h^{5/3}}
\end{equation}
where $M$ is the Manning coefficient, which we take to be $0.025$. (Typical values for
the Manning coefficient for a given substrate are empirically based. See \cite{Chow1959}
for a description and examples of values used in various applications.)

\newstuff{
Most tsunamis are generated by motion of the sea floor due to an
earthquake or submarine landslide, setting the entire water
column in motion.  The wave length is generally very long compared
to the depth of the ocean, and under these conditions the shallow water
equations \eqn{2dSWE} are generally appropriate.  This has been
confirmed in comparisons done by many groups (e.g.
\cite{YehLiuEtAl1995,KowalikKnightEtAl2005,%
LiuSynolakisEtAl2008,SynolakisBernardEtAl2007}), although in some
cases it is believed that dispersive terms may need to be included
(e.g.  \cite{GonzaleziKulikov:dispersion,Saito2010:dispersion}),
particularly when modeling tsunamis generated by submarine landslides,
which typically have short wavelengths (e.g.
\cite{LynettLiu2002,WattsGrilli:2003}).  In \Sec{apps} we illustrate
the use of GeoClaw for tsunami modeling using the shallow water
equations.  Adding dispersive terms would generally require the use
of implicit time stepping algorithms, which are not yet implemented
in GeoClaw. Development of an implicit version of the AMR
routines in Clawpack is a current project and this may be possible in the
future.  }

For other applications it is less clear that the classical shallow water equations are
sufficient.  For shallow flow on steep terrain, such as following a dam break for example,
vertical acceleration terms may need to be added to improve the model.  However, the
simple equations \eqn{2dSWE} are often still used for many practical
problems and can give fairly accurate
results.  In \Sec{malpasset} we display some dam-break
results from \cite{George2010}.
Some possible extensions to other depth-averaged systems of equations
are mentioned in \Sec{conclusions}.


\section{Numerical methods}\label{sec:num}

The algorithms used in GeoClaw are described in detail elsewhere;
\newstuff{see in
particular \cite{mjb-dg-rjl:actanum2011}.}   Here we
only give a brief summary with pointers to other sources for further reading.
GeoClaw is based on Clawpack, which
provides a general implementation of ``wave-propagation algorithms'', a
class of high-resolution finite volume methods in which each grid cell is viewed
as a volume over which cell averages of the solution variables $q$ are computed.
Logically rectangular grids are used and $Q_{ij}^n$ denotes the cell average in
cell $(i,j)$ at time $t_n$. In each time step the cell averages are updated by
waves propagating into the grid cell from each cell edge. These are Godunov-type
methods in which the waves are computed by solving a ``Riemann problem'' at each
cell edge. The Riemann problem is an initial value problem using the shallow
water equations together with piecewise constant data determined by the
cell averages of the dependent variables and topography on each side of the
interface.
\newstuff{The advantage of Godunov-type methods is that they provide a
robust approach to solving problems with discontinuous solutions, in
particular shock waves that generally arise in the solution to nonlinear
hyperbolic equations. In the shallow water equations, shocks are ``hydraulic
jumps'' or ``bores'', as often arise in practical flow problems.  The
Riemann problem defined at each cell interface allows for shock waves in
each Riemann solution and ``approximate Riemann solvers'' are used that
rapidly produce robust solutions as a building block for the numerical
method.  Correction terms are also incorporated so that the computed
solution is second-order accurate in smooth regions of the flow.}

GeoClaw employs a variant of the f-wave formulation, described in
\cite{George2008}, which allows the topography source terms to be directly
incorporated into the Riemann problem. The f-wave formulation of the
wave-propagation algorithms was originally presented for the shallow water
equations in \cite{db-rjl-sm-jr:vcflux}. In this approach it is the difference
in the {\em flux} normal to the cell interface that is split into propagating
waves (rather than the jump in $Q$), but only after modifying the momentum
components of the flux difference by the topography source terms. This is done
in such a way that the methods are well-balanced for the ocean at rest: if the
initial data is in equilibrium with zero velocities and a flat surface ($h+B =$
constant) then the modified flux difference is the zero vector, leading to
zero-strength f-waves and no change to the solution. This approach to well
balancing is discussed in more detail in \cite{db-rjl-sm-jr:vcflux,
rjl:wbfwave10, mjb-dg-rjl:actanum2011}  and Section 17.14 of \cite{rjl:fvmhp}.

The f-waves modify the cell averages on each side of the interface.
We also solve a ``transverse Riemann problem'' in which the waves moving normal to the
cell edge are split in the transverse direction and modify the cell averages in adjacent
rows of grid cells.  This improves stability and accuracy of the method and this general
approach is discussed in more detail in \cite{rjl:wpalg} and
Chapters~20--21 of \cite{rjl:fvmhp}, for example.
The f-waves are also used to define
correction terms modeling second derivatives normal to the interface that,
together with the transverse terms, make the method second order accurate for smooth
solutions.  Before calculating these terms, however, wave limiters are applied to the
f-waves to reduce their amplitude in regions where the solution varies rapidly.
This results in a ``high-resolution method'' that avoids nonphysical oscillations in
regions where the solution is rapidly varying.  This methodology has been well developed
in the context of shock-capturing for general nonlinear hyperbolic systems of equations
and leads to a robust method.

Developing a Riemann solver that works robustly in the presence of dry states is
particularly challenging --- it must handle the case where one state in the Riemann
problem is already dry as well as situations where a cell dries out as a wave
recedes, and must work robustly when the topography has arbitrary jumps from one
cell to the next.  Details of the solver we use are given in
\cite{George2006, George2008, George2010}.
\newstuff{
Incorporation of these dry state solvers is an important aspect
of GeoClaw, since we model the moving
shoreline or margin of a flow implicitly as
the interface between wet and dry cells.  This generally means that the
shoreline is represented by a stair-step pattern on a Cartesian grid.  By
using adaptive refinement we are able to use fine enough grids in regions of
interest that this can provide sufficient resolution, but this also means
that the dry-state algorithms must also function well in conjunction with
AMR grids and at the interfaces between grids at different levels of
refinement.  This was one of the more difficult aspects of
developing and debugging the GeoClaw extension.
}

The topography source terms (those involving $B_x$ and $B_y$ on the right
hand side of \eqn{2dSWE}) are incorporated in the Riemann solver
in order to obtain a well-balanced method and to handle the dry state problem.
On the other hand, the drag source terms are handled via a fractional step
approach: after each time step of the hyperbolic problem a time step is
taken in which the momenta are adjusted due to the drag terms. Coriolis
terms can also be incorporated in the same manner and this is included as an
option in GeoClaw, though for tsunami modeling at least this appears to be a
negligible effect, both in our own experiments and elsewhere in the
literature, e.g., \cite{KowalikKnightEtAl2005}.

In general Clawpack allows the solution of hyperbolic problems on any logically
rectangular grid, with an arbitrary mapping function specified that maps points in
computational space (a rectangular grid with uniform spacing) to the physical
domain.  In two space dimensions the grid cells are always assumed to be
quadrilaterals with linear cell edges joining vertices that are obtained by
applying the mapping function to the rectangular grid of vertices in the
computational domain.
\newstuff{On the sphere with longitude--latitude coordinates,
the cell edges are great-circle geodesics and the edge lengths and cell
areas must be measured on the sphere.}

The wave-propagation algorithms with transverse Riemann
solvers work very robustly, even on highly distorted grids or with non-smooth
mapping functions.  Unlike many approaches to mapped grids, we do not incorporate
metric terms that depend on derivatives of the mapping function into the differential
equations.  Instead, we always solve Riemann problems for the original set of
equations in the direction orthogonal to each cell edge.  The lengths of the edges
and the area of the quadrilateral cell then come into the formulas for updating
cell averages.  The transverse wave propagation also takes account of the fact that
adjacent cell edges are not necessarily orthogonal to one another.
For propagation on the earth, we use distances and areas as
measured on a sphere, although in principle this could be replaced by a geoid or
other surface.  Currently in GeoClaw the domain
choices are limited to the sphere with
longitude--latitude coordinates or purely Cartesian domains, primarily because the
more general routines for integrating topography data sets (see \Sec{topo})
over a general quadrilateral have not yet been developed.  Longitude--latitude
coordinates are suitable for modeling tsunamis on the earth, where
the domain of interest is bounded away from the poles.  In recent work we have also
explored another class of quadrilateral grids on the sphere and present some
results using AMR for a tsunami-type problem over synthetic bathymetry on
the whole sphere in \cite{mjb-dac-ch-rjl:amrsphere09}.

\section{Adaptive mesh refinement} \label{sec:amr}

In this section we describe the patch-based mesh refinement that is used to span
the orders of magnitude in spatial scales exhibited by many geophysical flow
problems. For example, to go from ocean scale propagation
to the resolution of small-scale coastline features in a single
tsunami model, meter-scale resolution may be needed in a small
subset of a domain that covers millions of square kilometers.
Multiple levels of patches can be used until a sufficiently fine
resolution is reached. We also give an overview of the numerical
algorithms needed to initialize and remove fine grid patches, and
present the organization of the time stepping procedures on the
grid hierarchy.

The time step on the refined patches is chosen
so that stability of the explicit finite volume
method is maintained.  This generally requires refining in time by the same
factor as in space.  For example, if the level 2 grids are refined in both $x$
and $y$ by a factor of 4 relative to level 1, then four time steps on all
level 2 grids must be taken for each time step on level 1.  The code is
organized so that the time step is first taken on level 1, which covers the
entire domain.  Then four time steps are taken on each of the level 2 grids.
In each time step it is necessary to fill in ``ghost cell'' values around
the edges of each level 2
grid in order to provide boundary conditions for the time
step.  For each ghost cell, the value is either taken from a neighboring
grid at the same level, if one exists, or otherwise is obtained by
space-time interpolation from the values on the underlying coarse grid,
which has already been advanced in time.

This same procedure is used recursively at all levels: after each time step on
the level 2 grids, the required number of time steps will be taken for
all level 3 grids and so on.  This AMR procedure is described in more detail
in \cite{mjb-rjl:amrclaw, mjb-dg-rjl:actanum2011}
and has been successfully used for many years in
Clawpack for problems such as shock wave propagation where dozens of grid
patches are used to track shock waves oblique to the grid.

The application to tsunami modeling prompted the addition of a new feature
to the code: the capability of specifying anisotropic refinement in time,
in which the time step from one level to the next
may be refined by a different factor than the
refinement in space.  This is crucial for problems where very fine grids are
used only near the coast of an ocean,
since in the shallow water equations the wave speed
is given by $\sqrt{gh}$.  In an ocean with a maximum depth of 4000 m, say,
this gives a wave speed of 200 m/s
(and in some regions the maximum ocean depth is much greater).
If a fine grid level covers only portions of the continental shelf with a
maximum depth of 100 m, say, then the maximum wave speed on this level is
roughly 32m/s.  Hence the refinement factor in time could be up to 6 times smaller
than the spatial refinement factor on this level.  Since the bulk
of the computational work is often on the finest grids, this can make a
substantial difference in computing time.

Grids are refined by flagging cells where the resolution is
insufficient and then clustering the flagged cells into refinement patches.
Flagging is done either using an error estimate (Richardson
extrapolation), by examining gradients of the solution (which will
detect where the largest waves are), or for the tsunami application simply by
flagging cells where the surface elevation is perturbed from sea level above
a specified level.  In addition, it is possible to specify rectangular
regions where a certain level of resolution is required.  This can
be used to insure that particular portions of a coastline are always refined
to a resolution of meters whereas the deep ocean uses a mesh width
of many kilometers. These routines are all part of the GeoClaw software,
controlled by user-specified tolerances along with  optional
user-specified regions in space and time where a minimum and maximum
allowable level refinement is specified.  Regardless of which method
(or combination of methods) is used to control the refinement
criteria, the result is a set of flagged cells needing to be covered by
finer grids.

Using heuristics from the pattern recognition literature \cite{mjb-rig:cluster},
these flagged cells are clustered into grid patches that are {\em
efficient}, in the sense that the grids do not contain too many unflagged
cells (which would be wasteful), while also not introducing too many
separate patches
(since there is boundary overhead associated with each fine grid
patch). The grids also should obey a {\em proper nesting} criterion
---  a grid patch at level 4 should be surrounded by a level 3 patch,
and not border directly on level 2 patches. Figures \ref{fig:ocean1}
and \ref{fig:island1}
show several frames from a 5 level
computation. In this example the grids were refined by a factor of
4 in going from each level to the next (hence a total refinement factor of
$4^4 = 256$ in each direction from coarsest to finest grids).

Every few timesteps the features in the solution needing refinement
will have moved, and the grid patches should move too. The grids
do not actually {\em move}; rather, at discrete times new grid patches are
created and their solution is interpolated from the finest previously
existing grids, which are then removed.  This interpolation step
must be done carefully. For example, a constant sea level should
be maintained even in the presence of variable bathymetry so that
no waves are generated solely from grid refinement.  This is accomplished by
interpolating the surface elevation $h + B$ for coarser grids
and then computing the depth $h$ in the fine cells by subtracting the
fine cell value of topography $B$.  This maintains conservation of mass
provided that the fine and coarse topography are {\em consistent}, in the sense
that the topography value used in a coarse cell should be the average of the
values in all fine cells that cover this cell.  This is ensured by
computing exact integrals of a single piecewise bilinear representation of
the topography, as described further in the next section.

Similarly when
a grid is removed, the coarse grid solution underneath it should
be the volume-weighted average of the fine grid cells it contained
so that mass is not lost or gained.  Most of the time these numerical
procedures are straightforward, but there are difficulties associated
with the wet-dry interfaces.
A coarse cell that covers a shoreline region will be either wet or dry,
depending on the level of the averaged topography.  Suppose it is dry, for
example.  When the cell is refined, typically some of the fine cells
will have to be initialized with
nonzero depth in order to represent the shoreline and to
maintain the constant sea level required before a wave arrives.  Hence it is
essential that water be introduced (mass increased) in this situation.
These subtleties are described more fully in \cite{mjb-dg-rjl:actanum2011}.

\section{Topography data sets} \label{sec:topo}
To use GeoClaw the user must provide one or more files that specify the
topography  for the terrain on which the flow evolves.
Each topography data set specifies the $z$ coordinate (relative to sea level,
for example) at a set of points on a rectangular grid (a longitude--latitude
grid if this coordinate system is being used).
Several different formats are allowed (see the documentation
\cite{geoclaw-doc}).

Appropriate data sets
for many regions of the earth are available online, for example from the
National Geophysical Data Center (NGDC) \cite{ngdc-grid}.  A few data sets for test
problems are
available in the GeoClaw topographic database \cite{geoclaw-topo}, and more
will be added in the future.
The test problems in GeoClaw include Python scripts to automatically
download this data as needed.
Some example instead use synthetic data (for example the tsunami model
presented in \Sec{tsunami}) and again a Python script is provided to create
this.

Some applications also require a dataset that describes the motion of the
topography relative to an initial topography, for example if seafloor motion
resulting from an earthquake or submarine landslide is used to generate a
tsunami.  In this case one or more files must also be provided to specify
the relative displacement at one or more times.


Often more than one topography file is used at different resolutions.  For
example, in a tsunami simulation a large region of the ocean may be modeled,
for which a fairly coarse resolution such as the 10-minute or 4-minute ETOPO2 data
available from NGDC is sufficient.  These have resolutions of roughly 18.5 km
or 7.5 km respectively
in each direction near the equator.  Since the wavelength of tsunami waves
is typically 10s to 100s of km this is sufficient for modeling propagation
across the ocean.  However, it is not sufficient for modeling inundation of
specific regions along the coast, and so this data must generally be
supplemented with one or topography files at much higher resolution over
small regions.

In GeoClaw, an arbitrary number of topography files can be provided for a
single run and at each point in space the topography will be determined from
the dataset covering this point at the finest resolution.  The user should
be aware, however, that this means there will generally be discontinuities
in the effective topography along the boundaries of fine scale datasets.

In the same way that $h_{ij}^n$ represents
the cell average of the fluid depth for a finite volume method,
we also need a cell average $B_{ij}$
of the topography in each grid cell.  Topography data sets generally give
the pointwise value of $B(x,y)$ on a grid of spatial locations.  To convert this
into cell averages, we construct a piecewise bilinear function that
interpolates the pointwise values and then compute the exact integral of
this interpolating function over a grid cell
to obtain $B_{ij}$.  This is easy to do if there is a single best-resolution
data set in the region around a cell, but if a grid cell covers an area
where two or more different data sets must be sampled then
computation of the integral is more difficult.   This often happens in realistic
tsunami simulations.  Fine grid topography for a small region of the coast may lie
entirely within one grid cell on the coarsest computational grids, for example.
We have implemented this quadrature in full generality
to guarantee that the topography values used are consistent between
different adaptive mesh refinement levels.

\section{Software tools and user interface} \label{sec:software}

\newstuff{
GeoClaw is comprised of a set of library routines written in Fortran 77 and 95,
in addition to a set of Python modules called PyClaw.  The Fortran library builds on the AMRClaw library of Clawpack, which was developed to apply AMR more generally to hyperbolic problems.
GeoClaw replaces many of the routines in AMRClaw with new ones specifically designed for geophysical flow problems.  Most of the core computation is done in the Fortran routines.  Python is used to fetch and operate on the topography,
setup the simulation run parameters, setup the plotting options and create plots.  More details about many of the tools mentioned in this section
can be found in the on-line GeoClaw documentation \cite{geoclaw-doc}.

We use the Subversion version control software and the Trac interface
as a development wiki and for its ticket system for bug tracking.
These can be found via the Clawpack webpage \cite{claw.org.url}.


\subsection{Problem specification} \label{sec:spec}

GeoClaw uses a Python script to prescribe most of the input parameters.  The
script constructs a data object that contains  values for all parameters that
GeoClaw needs to run,
and  then writes them out to a set of ASCII files that are read into the GeoClaw Fortran code at run time.
\newstuff{
There are several reasons for taking this approach to prescribing the input
files.
A major advantage is that it is easier to maintain
backward compatibility as Clawpack and GeoClaw evolve. If a new feature is
added that requires new input parameters, these can be added to the Fortran
code and default values added to PyClaw so that old applications continue to
run without change to the user's input script.
It is also easier to write a flexible parser in Python than
in Fortran, and the use of a Python script for setting the parameters allows
the user to use loops or functions, for example to define an
array of desired output times using the {\tt
linspace} command of {\tt NumPy}.
Sample input files can be viewed in the documentation (see the sections on
{\tt setrun.py} or the
sample codes to accompany this paper at \cite{awr10-web}.

The Fortran code is a stand-alone code that reads the input files
created by the Python script as data and handles memory allocation using a
combination of Fortran 95 dynamic memory operations and a large work array
that is managed by our own Fortran routines to efficiently allocate and
deallocate storage needed for all of the AMR grids.  The Fortran {\tt
allocate} statement is used only if the size of the
work array needs to be increased during the computation, in which case it is
generally doubled in size.
Software such as {\tt f2py} allows one to easily call Fortran from Python
code and a future project is the ability to control time-stepping from a
Python wrapper that would be able to produce plots as the computation
proceeds, for example.  In one space dimension the PyClaw software also
includes a pure Python version of the finite volume methods
with no Fortran component.  This is useful as a test bed and teaching tool
but runs considerably slower than the Fortran version.
}

\subsection{Plotting} \label{sec:plotting}
}
Early versions of the Clawpack software included a
set of Matlab plotting routines for visualizing the results.  Some specialized
versions of the Matlab plotting routines were created for dealing with
topographic data sets and are available in GeoClaw.   Recently, however,
the main development of plotting tools for Clawpack has shifted from Matlab
to Python for a number of reasons.  Many users of Clawpack do not have
access to Matlab and it is desirable to have an open source alternative.  Moreover,
in the past few years substantial improvements have been made in Python
plotting packages that provide quality that equals or exceeds that of Matlab
graphics.  For two-dimensional plots of the type shown in this paper, we use
the {\tt matplotlib} module \cite{matplotlib}.

We have developed a Python plotting module that allows the user to easily
specify a set of plots to be produced for each frame of a simulation.
When AMR is being used it is necessary to loop over all grids and combine
the solution on each grid into a single plot.  It is often desirable to
combine several plots in a single figure.
For example, we may want to do a
pseudo-color plot of the water surface elevation using one color map while the
topography in dry regions is also plotted with a different color map.
Contour lines of bathymetry may be added to this along with indications of
locations of tide gauges, resulting in a plot such as the ones shown in
\Sec{apps}.
The logic of looping over the grids is handled by the plotting module and the
interface provides a mechanism for specifying a variety of different plots or
combinations of plots on a single axis without the user needing to deal with the
AMR data structures.  Other useful tools such as codes for dealing
with the topographic data sets and colormaps appropriate for these problems are also
included.

There are also several ways the user can view plots coming from a
simulation.  There is an interactive Python module {\tt Iplotclaw} for stepping
through the frames of a simulation and producing the plots on the screen,
facilitating data exploration via zooming in on features of interest, for
example. Alternatively, it is easy to generate a set of hardcopy files in
formats such as {\tt png} or {\tt jpg},
one for each figure at each time frame, together
with a set of webpages designed to easily browse through the collection of
plots.  Webpages are automatically created to loop through all frames of
each figure, creating an animation that is often extremely useful in
developing a better understanding of the time-evolution of the flow.  This
set of webpages also simplifies the process of archiving past experiments
for later viewing, or for sharing sharing simulations with others.
Examples can be viewed on the webpages that accompany this paper
\cite{awr10-web} and in the gallery of Clawpack and GeoClaw applications
in the documentation.
Many of these tools have been developed with the aim of
encouraging users to adopt the paradigm of {\em reproducible
research} in computational science. The approach we have taken with
Clawpack is discussed in more detail in \cite{rjl:cise09}.

For three-dimensional
surface plots we are currently investigating several options, including
Mayavi \cite{mayavi}, which is included (along with {\tt matplotlib}) in the
Enthought Python Distribution \cite{epd},  and VisIt, an open
source visualization package being developed at Lawrence Livermore National
Laboratory \cite{llnl:visit}.  VisIt provides more functionality for
large scale visualization problems and is designed to work well with AMR
data and distributed memory supercomputers.  Development of 3D plotting
tools for GeoClaw and Clawpack more generally is an on-going project.

\newstuff{
\subsection{Extending GeoClaw} \label{sec:extending}

An important aspect of GeoClaw is the ease at which GeoClaw can be
extended to include other physics and algorithms.  These extensions
can be added in a number of ways, for example
through modification of the Riemann solvers
or by adding a source term.
Different physics can be incorporated into the Riemann solver, to model
problems for which the shallow water equations are not sufficient.
\newstuff{
Source terms,
represented by $\psi$ on the right hand side
of equation \eqref{2deqn}, are used to incorporate
bottom friction terms and Coriolis terms, for example.
These can be extended to model
other terms, such as the wind forcing of a tropical storm to model storm
surge.
A fractional step procedure is used in which time steps on the homogeneous
hyperbolic system are alternated with time steps on the source terms, and so
the user need only supply a subroutine that takes a time step on the system $q_t
= \psi(q)$.
}

Another aspect that users may way want to modify is the algorithm
used to flag grid cells for refinement.
Currently, GeoClaw uses displacement from sea level
in addition to a set of fixed refinement regions that can
be specified by the user.  One could alternatively refine based on
the momentum or speed of the fluid, for example.

The Clawpack and GeoClaw documentation contains
more information on these routines and how they can be extended.
A number of other extensions are currently being developed and some of these
are briefly discussed in \Sec{conclusions}.
}

\section{Applications} \label{sec:apps}
We briefly describe three applications of GeoClaw.  The first is a new
synthetic tsunami test problem.
The second example models the 27 February 2010 earthquake in Chile as an
illustration of the use of real data sets, and is included in the GeoClaw
distribution.  Variants of these problems are also presented in
\cite{mjb-dg-rjl:actanum2011}.
The third example is the simulation of the Malpasset dam catastrophe of
1959, which has been well studied and often used as a benchmark problem, and
for this problem we
give a summary of the GeoClaw results that were first presented in
\cite{George2010}.


\subsection{Synthetic tsunami test}\label{sec:tsunami}
\begin{figure*}
\hfil\includegraphics[width=2.1in]{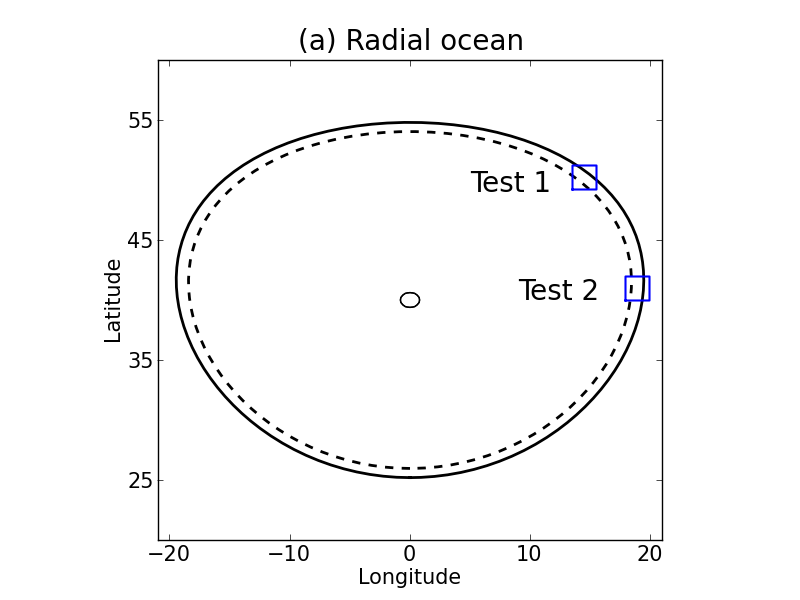}\hfil
\hfil\includegraphics[width=2.1in]{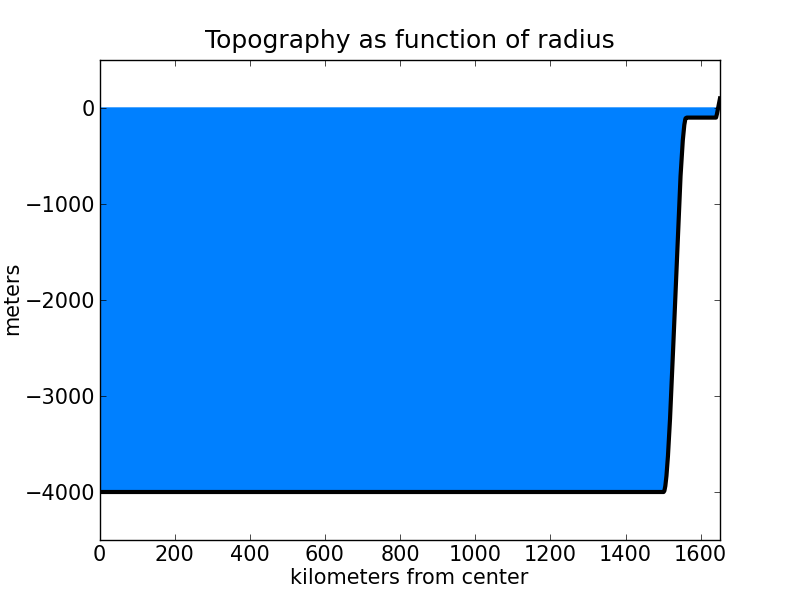}\hfil
\hfil\includegraphics[width=2.1in]{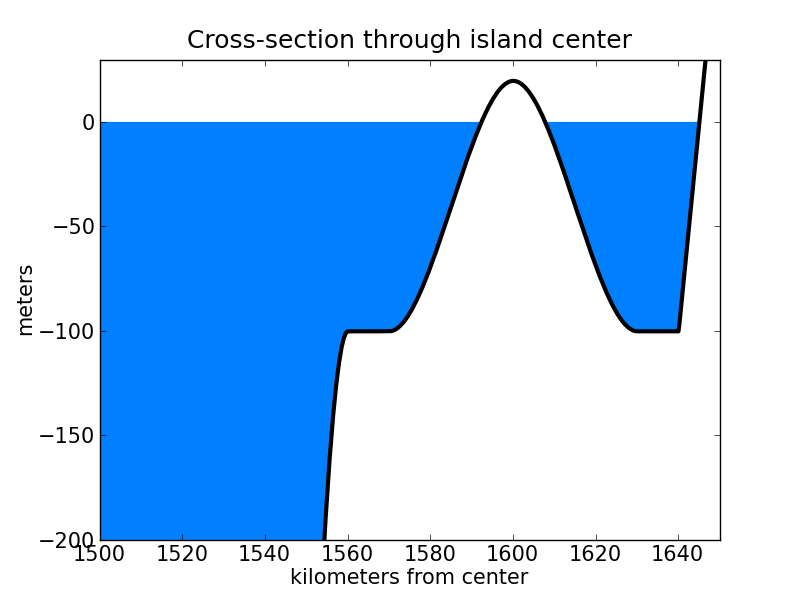}\hfil
\caption{\label{fig:radocean}
(a) Geometry of the radially symmetric ocean.  The outer solid curve
is the position of the shoreline, with constant distance from the
center when measured on the surface of a sphere.  The dashed line
shows the extent of the continental shelf.
The boxes labelled Test 1 and Test 2 are regions where an island is located
in the tests presented in the following figures.
The small circle near the center shows the extent of the hump of water used
as initial data.
(b) The topography defined by the piecewise cubic function \eqn{Bocean}.
(c) A zoom view of the topography of the continental shelf along
the ray going throug the center of the island.
  }
\end{figure*}

First we present some results obtained using a synthetic data set
that has been designed to illustrate the power of our adaptive
refinement approach with a realistic range of spatial scales, but
in a context where it is also possible to assess the accuracy of
the solution. We start with a radially symmetric ocean that has a
depth depending only on distance  from some central point, which
we take to be longitude $x_0=0$ and latitude $y_0=40^\circ$N (see
\Fig{radocean}(a)).
Distance is measured as the great circle distance on a spherical
earth by the formula
\[
\begin{split}
&d(x,y;~ x_0,y_0) = 2 R \arcsin((\sin(0.5(y-y_0))^2 \\
&\qquad\qquad\null + \cos(y_0)\cos(y)\sin(0.5(x-x_0))^2)^{1/2}).
\end{split}
\]
where $R=6367.5\times 10^6$m is the average radius of the earth.
In this formula we assume the longitude-latitude pairs
$(x,y)$ and $(x_0,y_0)$ are in radians to avoid the factors $\pi/180$.
The bathymetry profile is shown in \Fig{radocean}(b), where
the horizontal scale is in kilometers and the vertical scale in meters. The central
portion of the ocean is flat and is bounded by a
continental slope and flat continental shelf, followed by a linear beach.
We use a continuous piecewise cubic function whose derivative is also continuous except at
$r_3$, the start of the beach:
\eql{Bocean}
B(d) = \bpwdef z_1 \when d \leq r_1\\
C(d)
\when r_1\leq d\leq r_2\\
z_2\when r_2\leq d\leq r_3\\
z_2+\sigma(r-r_3) \when d \geq r_3.
\epwdef
\end{equation}
where the cubic $C(d)$ is given by
\[
C(d) =
z_1 + \frac{(x_2-z_1)(d-r_1)}{r_2-r_1}\left( 1-2\frac{d-r_2}{r_2-r_1}\right)
\]
and smoothly connects the ocean floor to the continental shelf.
Here $r_1=1500\times 10^3$m is the start of the continental
slope, $r_2=1580\times 10^3$m is the start of the flat continental
shelf, and $r_3 = 1640\times 10^3$m is the start of the beach, which
has slope $\sigma = 0.02$.  We take $z_1=-4000$m for the depth of
the ocean and $z_2=-100$m for the depth of the shelf.  The initial
shoreline is at $1645\times 10^3$m.

\begin{figure*}
\hfil\includegraphics[width=3.1in]{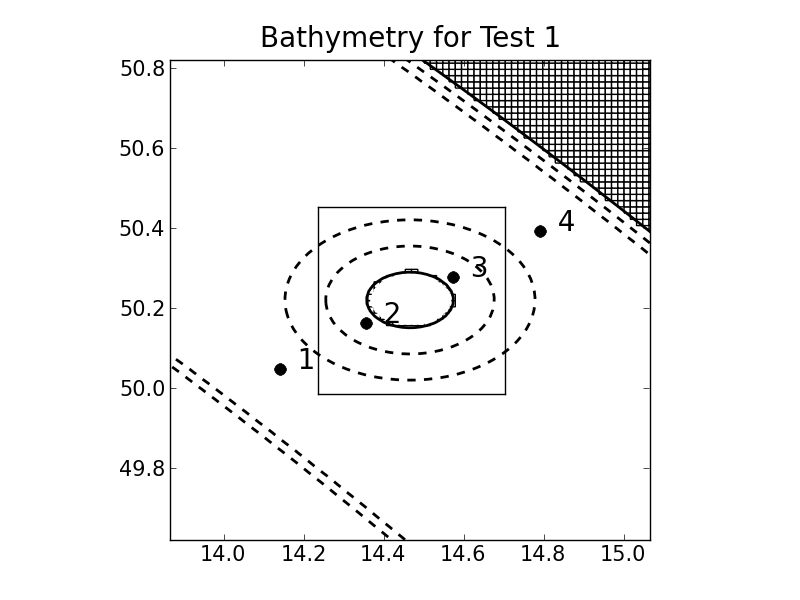}\hfil
\hfil\includegraphics[width=3.1in]{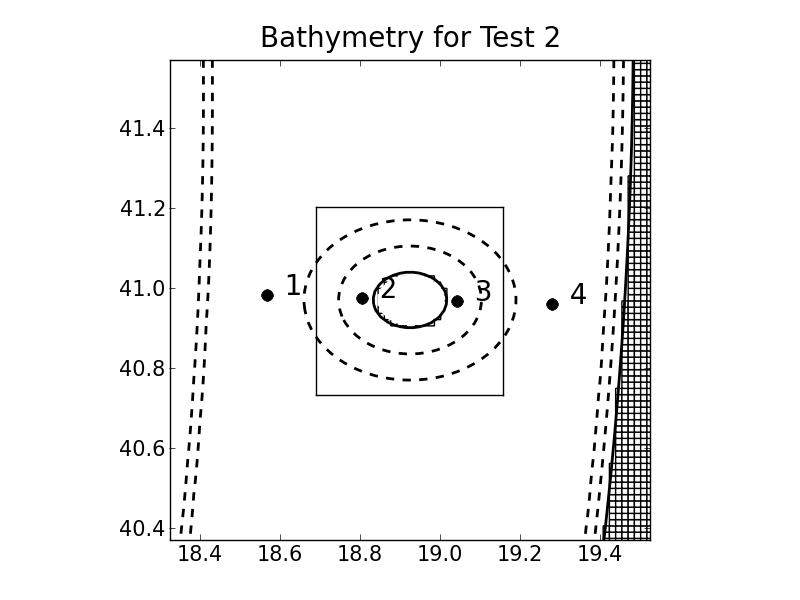}\hfil
\caption{\label{fig:islandbathy}
Topography in regions near the island for Test 1 and Test 2.
The solid contour lines are shoreline ($B=0$) and the dashed contour lines are at
elevations $B=-40,-80,-120,-160$m.  Note that the continental shelf has a uniform
depth of $-100$m.
The level 4 grid is shown onshore.
The rectangle around the island shows the level 5 grid in each computation,
which is refined by an additional factor of 4 in each direction from the
level 4 grids.  The location of four gauges is also shown.
The time history of the surface at these gauges is shown in Figure
\protect\ref{fig:gauges}.
  }
\end{figure*}

As a smaller scale feature we add an island on the continental shelf
at one point that can be varied.  The island is defined by a piecewise
function of distance about a specified center point $(x_1,y_1)$
(again using great circle distance).  The center point is chosen
to have distance 1600 km from the center of the ocean (hence 45 km
off shore).  The island is specified by
$B_2(d) =
120(1 - (d/r_4)^2(1 - 2(d-r_4)/r_4))$ for $d\leq r_4$ and zero outside this
radius.  Here
$d$ is the distance from $(x_1,y_1)$ and the radius of the island
is $r_4=30\times 10^3$m.  The island rises from the continental
shelf to a peak height of 20 m above sea level.  \Fig{radocean}(c)
shows how the cross section is modified along the radial slice that
passes through the center of the island.

The full bathymetry at any longitude--latitude point is thus given by
\[
B(x,y) = B_1(d(x,y;~ x_0,y_0) + B_2(d(x,y;~x_1,y_1))
\]
A Python script that can be used to generate bathymetry files with
arbitrary resolution is provided in the directory for this example, which
can be downloaded from \cite{awr10-web}.
\Fig{radocean}(a) shows the entire ocean in longitude--latitude coordinates.
The dashed line  indicates the extent of the continental shelf.
As initial data we take an ocean at rest and add a Gaussian hump
of water at the center of the ocean:
\[
\begin{split}
h(x,y,0) &= 20\exp(-0.5\times 10^{-9}d(x,y;~x_0,y_0)^2) \\
&\qquad\null - B(x,y).
\end{split}
\]
The innermost contour on \Fig{radocean}(a) shows the contour where
the initial hump has an elevation of 2m above sea level, 10\% of
its peak value.

\begin{figure*}
\hfil\includegraphics[width=3.1in]{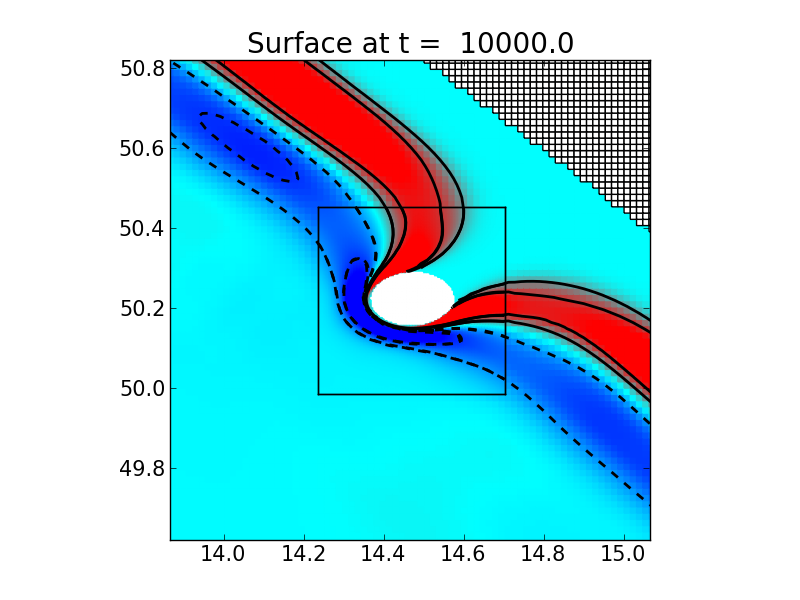}\hfil
\hfil\includegraphics[width=3.1in]{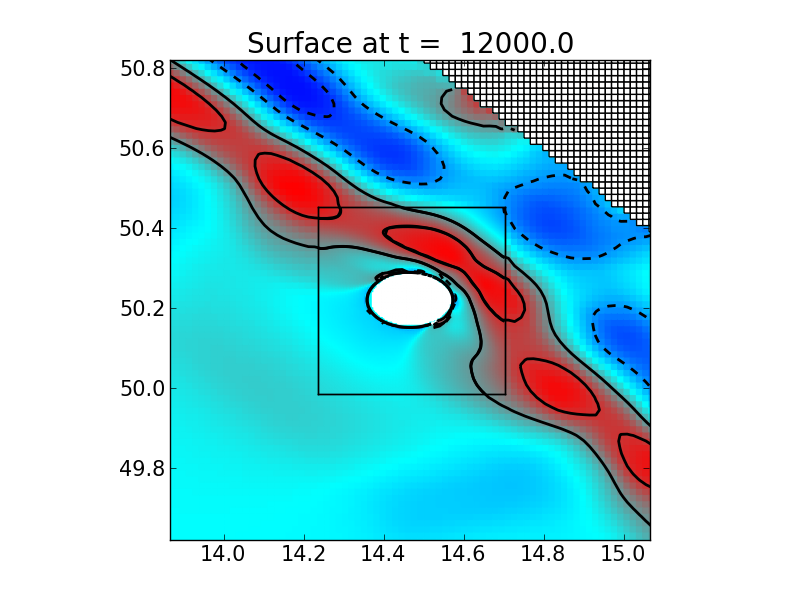}\hfil

\hfil\includegraphics[width=3.1in]{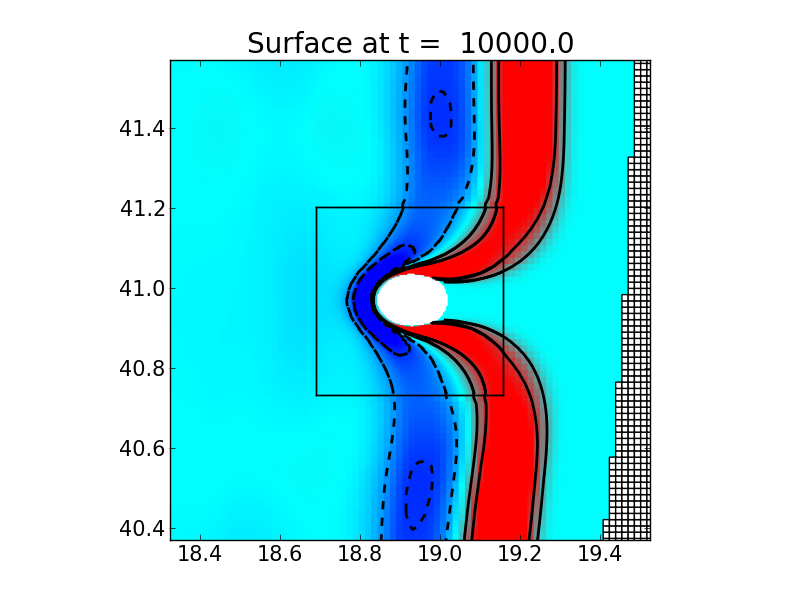}\hfil
\hfil\includegraphics[width=3.1in]{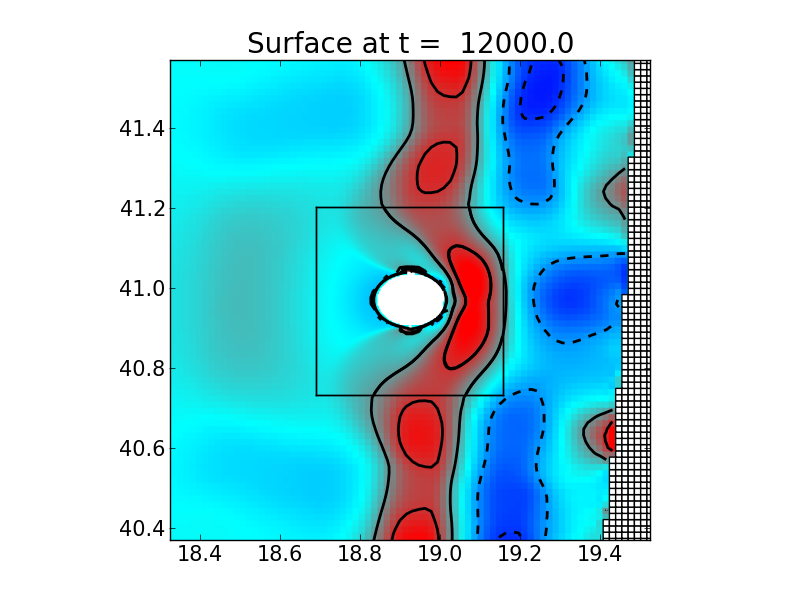}\hfil
\caption{\label{fig:island1}
Top row: Surface elevation at two times for Test 1.
Bottom row: Same times for Test 2.
At time $t=10000$ seconds the wave is approaching the shore.
At time $t=12000$ the wave has reflected off the shore and is outgoing.
In all cases the same contour levels are shown: the solid contours are at
elevations of 0.4m and 0.8m above sea level, the dashed contours are at the
same elevations below sea level.
The inner rectangle is the interface between level 4 and level 5 grids.
The level 4 grid is shown onshore.  The level 5 grid is finer by a factor of 4 in
each direction.
  }
\end{figure*}

We have performed several different runs in which the central point
of the island $(x_1,y_1)$ always has the same distance from $(x_0,y_0)$ but is
located at different angular locations.  The solution to the shallow
water equations with this set up should be exactly radially symmetric if
there were no island. With the island the solution with different
island locations should ideally be rotations of one another.  This
is a good test of the numerical method since the grid orientation
to the shoreline near the island varies greatly depending on the its location.
The two nearshore boxed regions in \Fig{radocean}(a) indicate
the two test cases considered here.  In each case an island is centered in
the 2-degree square box seen in \Fig{radocean}(a).  See \Fig{island1} for
closeups of these regions for the two tests.

The longitude--latitude domain
$[-20,20]\times [20,60]$ is covered with a coarse $40\times
40$ grid, so the mesh width is one degree on this level 1 grid.
Note that one degree
of latitude is about 111km and one degree longitude varies from
38km at $60^\circ$N to 96km at $20^\circ$N.  We use 5 levels of
mesh refinement, with refinement factor 4 in going from each level to
the next, and hence a total refinement factor of  $4^4=256$
from the coarsest to finest grids.
The level 5 grid has a mesh width $1/256 \approx .0039$
degrees, or 434m in the latitude direction.
(More levels or higher
refinement ratios could be used to refine further at particular
points along the shoreline.  In the tsunami computation presented
in \cite{LeVequeGeorge2008}, for example, we used a total factor of 4096 refinement
between coarsest and finest levels.)

Refinement to level 3 is allowed over the portion of the ocean that
is in the direction of the study area.  Refinement to levels 4 and
5 is only allowed in a small region near the island, as seen in
\Fig{island1}.

\begin{figure*}
\hfil\includegraphics[width=3.1in]{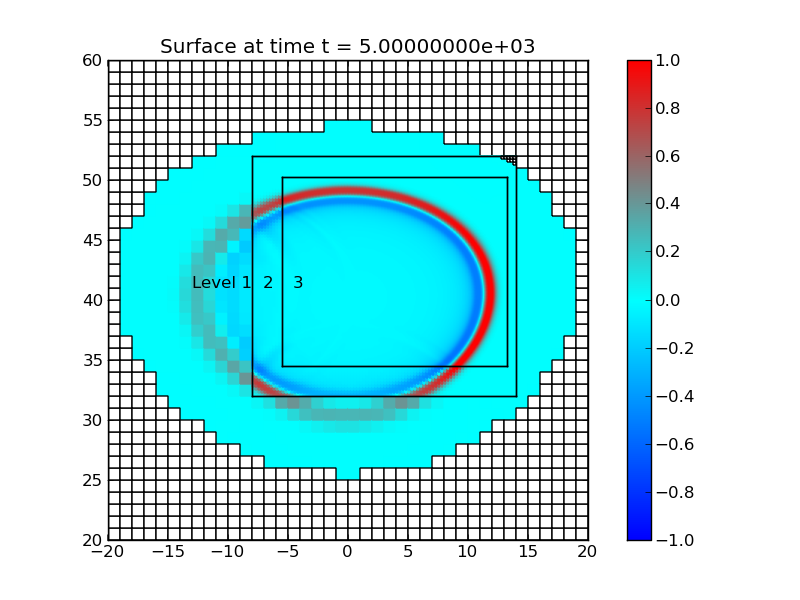}\hfil
\hfil\includegraphics[width=3.1in]{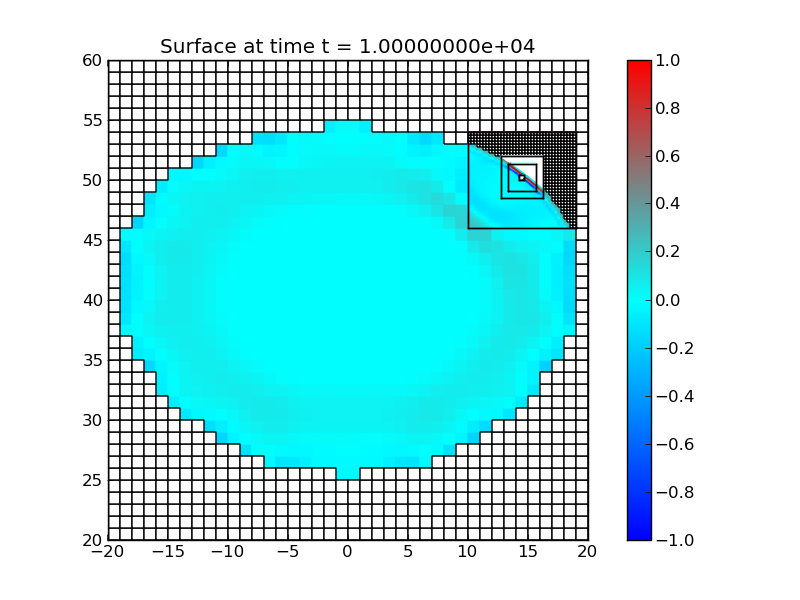}\hfil
\caption{\label{fig:ocean1}
Computed surface elevation for Test 1 at two different times.  Left:
at time $t=5000$ seconds, at which point at most
three refinement levels are allowed, and only in part of the domain.  Right:
at time $t=10000$, when 5 levels are allowed but only near the island, which is
not visible on this scale.
See Figure~\protect\ref{fig:island1} for a zoom of the region around the island.
  }
\end{figure*}

The Gaussian initial hump spreads out into a wave that propagates
radially.  \Fig{ocean1} shows the sea surface elevation at time
$t=5000$ seconds, for a calculation in which only 3 levels of
refinement have been allowed so far, in the case where the island
is located in the square indicated as Test 1 in \Fig{radocean}(a).  The
edges of refinement patches are drawn and the spreading wave is
poorly resolved on the coarse grid, but well resolved in the refined
regions.

%

\begin{figure*}
\hfil\includegraphics[width=3.1in]{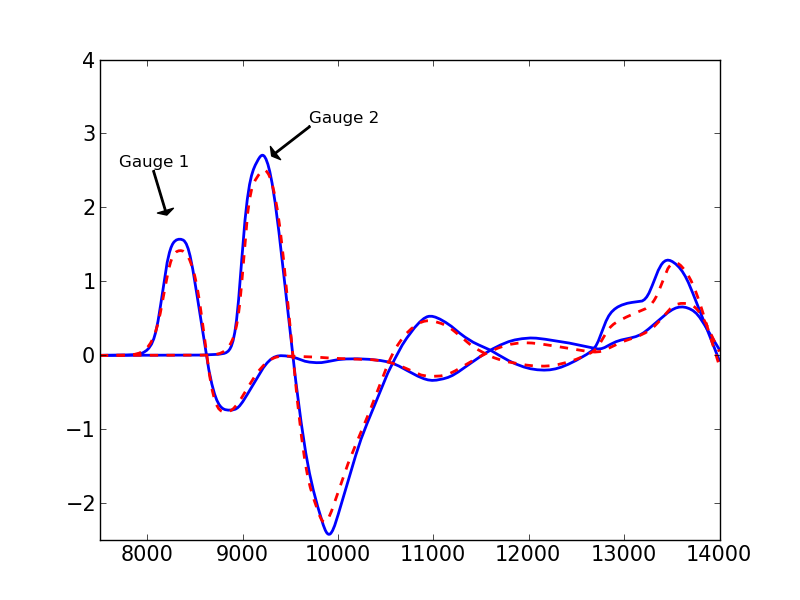}\hfil
\hfil\includegraphics[width=3.1in]{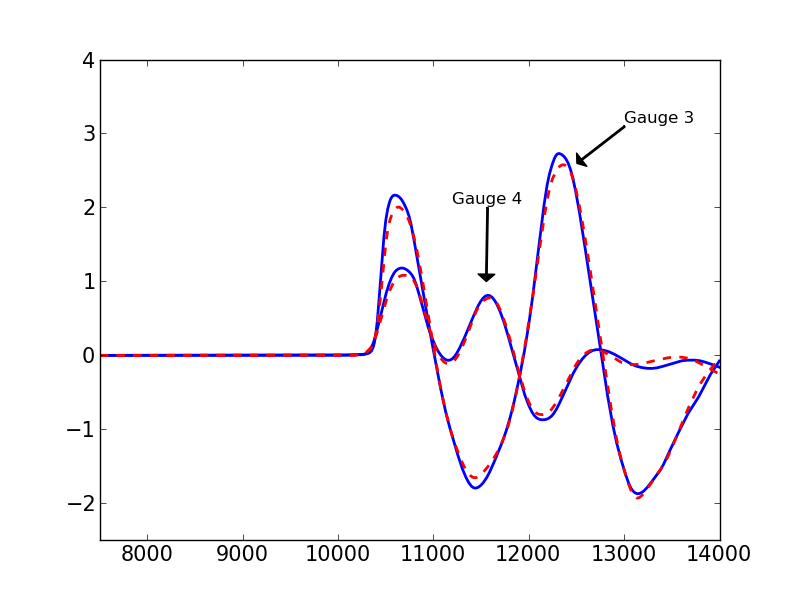}\hfil
\caption{\label{fig:gauges}
Comparison of gauge output from Test 1 and Test 2, showing the surface
elevation (vertical axis, (m)) as a function of time (horizontal axis, (s)) for the gauges shown in
Figure~\protect\ref{fig:islandbathy}.
In each case the solid blue curve is from Test 1 and the dashed red curve is
from Test 2.
  }
\end{figure*}

Note also that the calculation is done on the surface of the sphere
and so the wave spreads as a circle on the sphere, but in
longitude--latitude space the wave front is not circular.   The wave appears
to be on track to reach all points at the shore simultaneously, as should
happen, and this is confirmed in Figures~\ref{fig:island1} and
\ref{fig:gauges} which show nearly identical time histories at two locations
near the shore. The top row of \Fig{island1} shows three later times for an island
located in the box labelled Test 1 in \Fig{radocean}(a).
The bottom row of this figure shows the same three times for
a second test run, in which the island was located in the box labelled Test
2 in \Fig{radocean}(a).

\Fig{gauges} shows tide gauge data from the computation at four gauge
locations on the radial line from the center of the ocean that passes
through the center of the island.  Gauges 1 and 2 are distance 1570 km and
1590 km from the center, on the seaward side of the island, while Gauges 3
and 4 are distance 1610 km and 1630 km from the center of the ocean, on the
lee side. Gauges 1 and 4 are in regions that are never refined beyond level
4, while Gauges 2 and 3 are with the region refined to level 5.
Each figure shows two curves for each gauge, one from Test 1 (solid lines) and
one from Test 2 (dashed lines).
In principle these should lie on top of each other and in fact the
agreement is quite good.

Each of these calculations (for Test 1 and Test 2) required roughly
18 minutes of one processor on a 32 bit, 2.26GHz MacBook Pro laptop computer.
Approximately 55 million grid cells were advanced in time over the entire
computation, of which roughly 14M were on level 3, 28M on level 4, and 11M
on level 5. Levels 1 and 2 combined accounted for less than 1M.

\subsection{The 27 February 2010 tsunami}\label{sec:chile2}
The GeoClaw code has been used to model several historical tsunamis
using bathymetry and topography data sets obtained from
NDGC \cite{ngdc-grid} and other sources.
Some simulations of the 26 December 2004 tsunami in the
Indian Ocean following the Sumatra-Andaman earthquake are presented
in \cite{GeorgeLeVeque2006, LeVequeGeorge2008}
and several other studies are under way to be published elsewhere.
See \cite{awr10-web} or \cite{geoclaw-doc} for links to some animations.

Here we present some results for the 27 February 2010
Chile event.  This example is included in GeoClaw as a sample to illustrate
the use of topography data sets.  In this case we use 10-minute
ETOPO2 topography from NGDC \cite{ngdc-grid}.
The seafloor motion is generated using the
Okada model \cite{Okada1985}, which translates earthquake parameters taken
from \cite{usgs-chile2010} into
seafloor deformation, using a general
Python function implementing the Okada model that is included in GeoClaw.

\newstuff{
It should be noted that there are many uncertainties in the data used for
tsunami modeling.  In particular, the motion of the seafloor that generates
the tsunami is generally not well determined.  Even after seismologists use
a multitude of seismic signal measurements to perform source inversion
and determine the slip of the earthquake, this typically takes place
many km beneath the seafloor.  The seafloor displacement is dependent
on the subsurface geologic structure and is only approximated by
the Okada model, which assumes an isotropic material in a half-space.
One use of tsunami modeling is to perform source inversion directly
from measurements of the tsunami to estimate
directly the seafloor displacements.
This is the primary purpose of the DART buoys
(Deep-ocean Assessment and Reporting of Tsunamis),  and
other devises that
measure the pressure at the seafloor in deep water and
make possible the early estimation of a tsunami's magnitude and
destructive potential \cite{Meinig:2006p819}.
However, the paucity of such data makes it difficult
to obtain detailed reconstructions of the seafloor displacement.
}

\begin{figure*}
\hfil\includegraphics[width=3.1in]{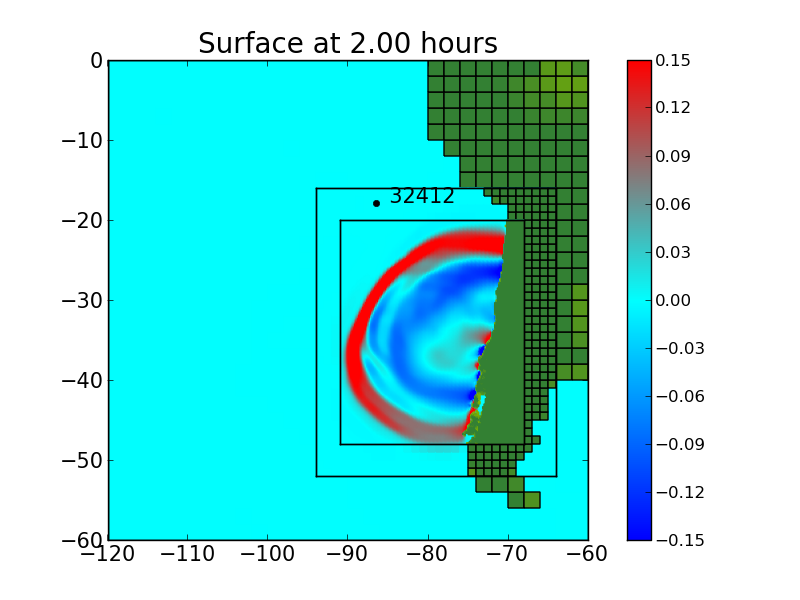}\hfil
\hfil\includegraphics[width=3.1in]{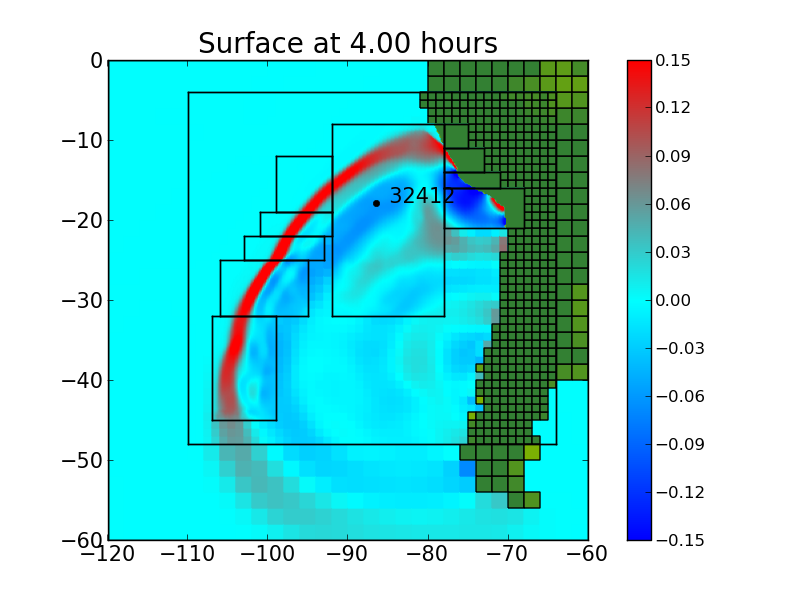}\hfil
\vskip 5pt
\hfil\includegraphics[width=3.1in]{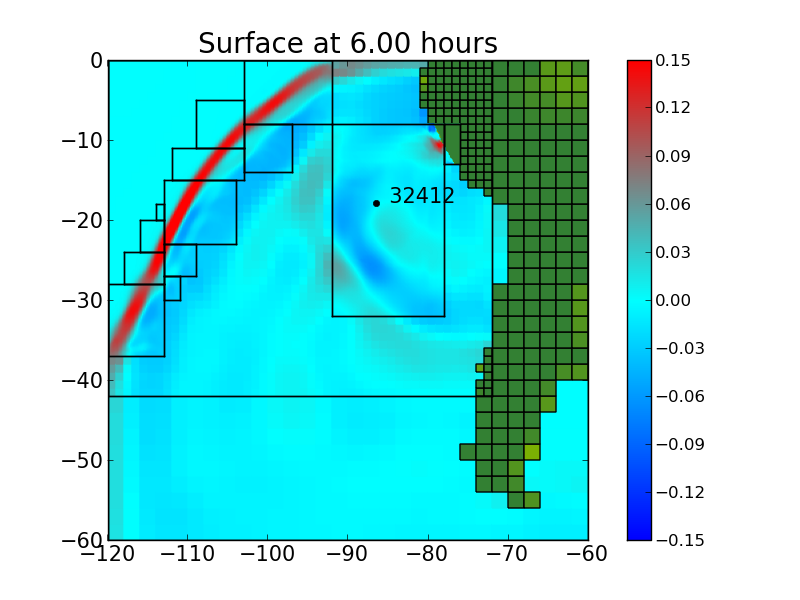}\hfil
\hfil\includegraphics[width=3.1in]{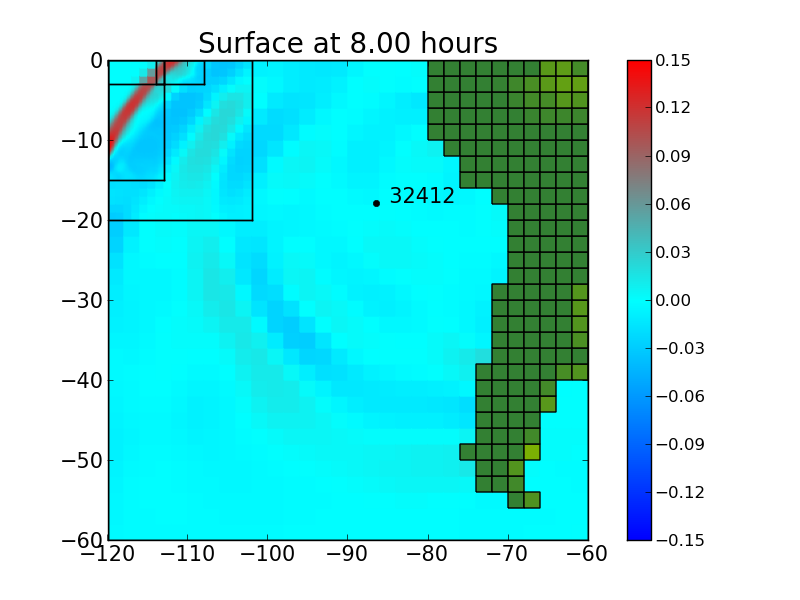}\hfil
\caption{\label{fig:chile}
Four frames from a 3 level simulation of the 27 February 2010 Chile event.
The location of DART buoy 32412 is also indicated, for which the time
history is shown in \protect\Fig{dart}.  Note the reflection from the
Gallapagos near the equator at 6 hours.
  }
\end{figure*}

\begin{figure*}
\hfil\includegraphics[width=6.0in]{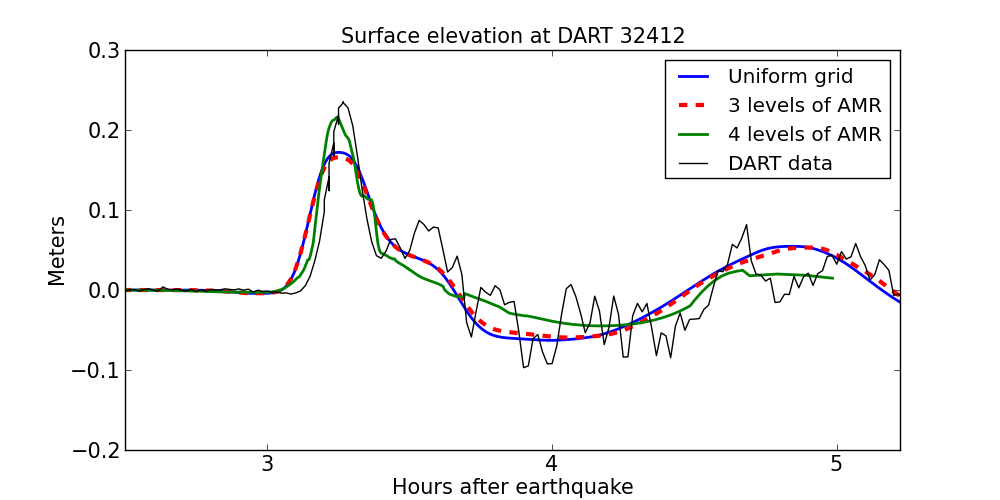}\hfil
\caption{\label{fig:dart}
Time history at DART buoy 32412, together with computational results
from three GeoClaw runs as described in the text.
  }
\end{figure*}

\Fig{chile} shows four frames from a 3 level simulation.  Level 1 has cell
size 2 degrees. Refinement factors of 2 and 6 are used, so the finest
grid has cell size of $10'$ and matches the topography data.
The dot labelled 32412 shows the position of DART buoy 32412
\cite{dart-32412}, which collected data during the event.
\newstuff{
\Fig{dart} shows this data together some GeoClaw modeling results.
The 3 level simulation is as described above.  The uniform grid simulation
was performed on a $360\times 360$ grid with $10'$ resolution, corresponding
to the finest level of the 3 level run.  The 3 level results lie on top of
the uniform grid results at early times, as one would hope to see.  At later
times the 3 level simulation is less accurate because the region near the
DART buoy is no longer refined once the main tsunami wave has moved past it.
\Fig{dart} also shows results obtained with a 4 level simulation using
refinement factors 2,6, and 8, so that the finest has $1.25'$ resolution.
With this resolution the leading peak is captured better and the amplitude
of the primary wave is well estimated.
}

Note that there is very little evidence
of spurious reflected waves at the refinement boundaries in this figure
(or in the figures from the previous example).
This is true in general with the AMR approach used in Clawpack.  Moreover,
for problems where the computational domain does not cover the full
ocean (such as in \Fig{chile}),  it is important that the method does not
generate spurious numerical reflections at these outflow boundaries.
The Godunov-type wave-propagation algorithms do a very good job of
providing non-reflecting boundary conditions simply by using
constant extrapolation into ghost cells at the domain boundaries: the values
in interior cells adjacent to the boundaries are copied into ghost cells.
Then solving the Riemann problems at the interfaces along the boundaries
results in zero-strength waves propagating into the domain, and hence no
apparent reflection of the out-going waves.

\newstuff{The 3-level
computation ran in about 1.5 minutes on a 64 bit, 2.26GHz MacBook Pro laptop,
advancing
21 million grid cells.
By contrast, the $360\times 360$ uniform grid computation
on this domain (at the resolution of the finest AMR grid)
required about 8 minutes of computer time, advancing 137 million grid cells.
}

The advantage of
AMR is clear, even for this problem where we are not zooming in on regions
of the coast to model inundation.

The uniform grid calculation exhibits a larger number of ``cells advanced
per second of computation'' \newstuff{(285K vs.\ 233K)}, due to the overhead of
adaptive refinement.  This overhead is greater in GeoClaw than normally found in
Clawpack because of the need to recompute the topography in each grid cell
each regridding time.  In this calculation we regrid every 3 time steps to
insure waves do not leave refinement patches between regridding.
In the future we hope to improve the efficiency of regridding the
topography.

\subsection{Riverine and overland flooding} \label{sec:malpasset}
The shallow water equations are often used to model riverine or overland
flooding problems, such as those due to dam/levee breaches (e.g.
\cite{HervouetPetitjean1999,ValianiCaleffiEtAl2002,Alcrudo1999a}). For flooding
problems in rugged mountainous terrain, where rapidly varying contours in the
topography create highly irregular domains, adaptive mesh refinement can be a
valuable tool because the optimal grid resolution is highly spatially and
temporally dependent yet unpredictable before doing the computation.
A common approach for modeling floods in
complicated topographic regions is to use static irregular meshes that are fit
to the topography in some fashion (e.g. \cite{ValianiCaleffiEtAl2002}).
However, by using adaptive mesh refinement, uniform rectangular grids can be
used for such problems, resolving the flood on an evolving patchwork of finer
grids that advance with the flood waves through topography.  This makes it much
simpler to set up a general problem.

\begin{figure*}
\begin{tabular}{c c c}
\includegraphics[width=4.8cm]{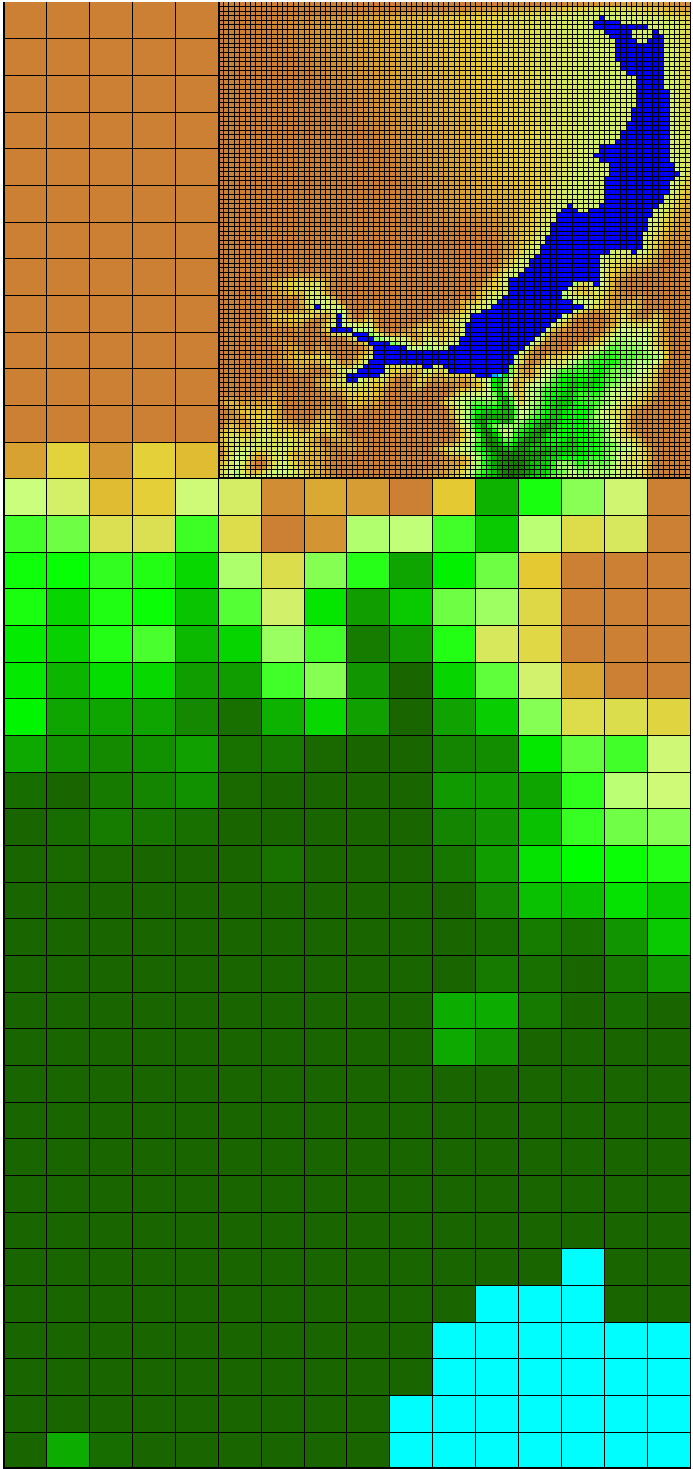} &
\includegraphics[width=4.8cm]{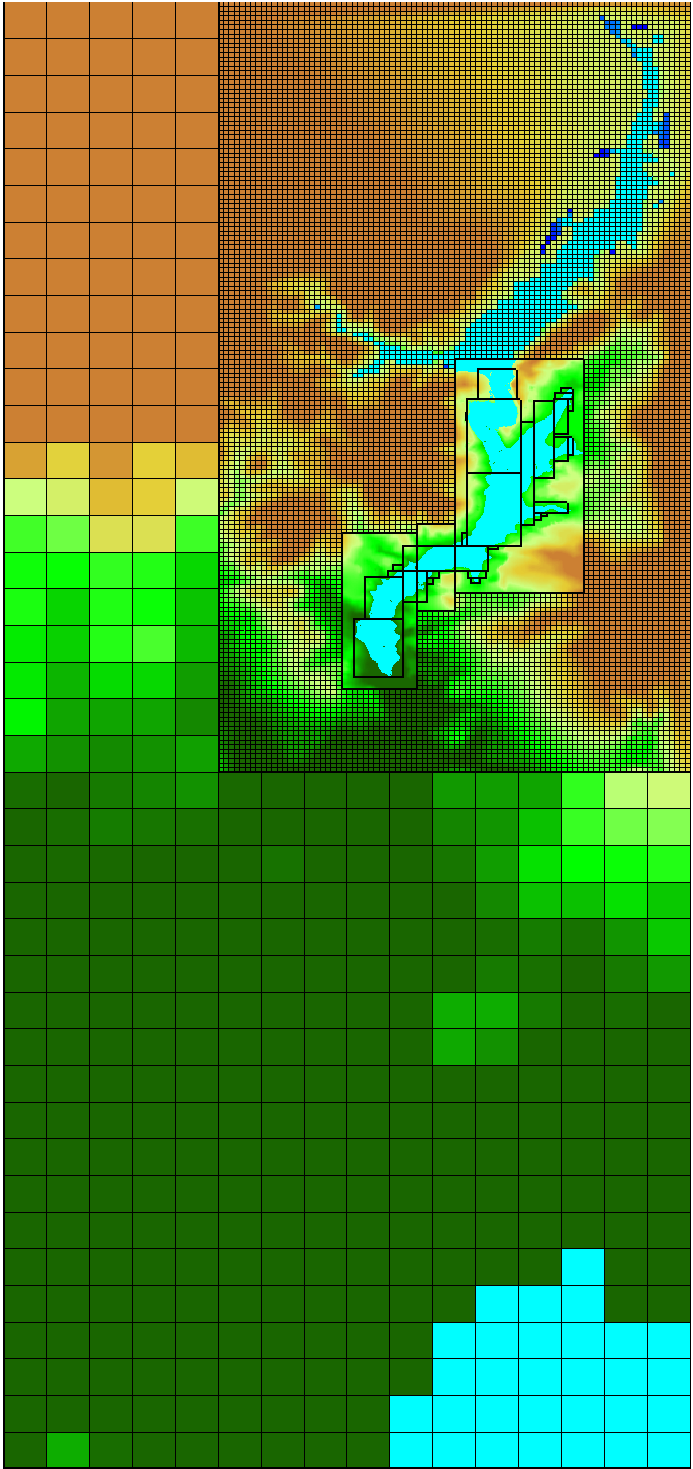} &
\includegraphics[width=4.8cm]{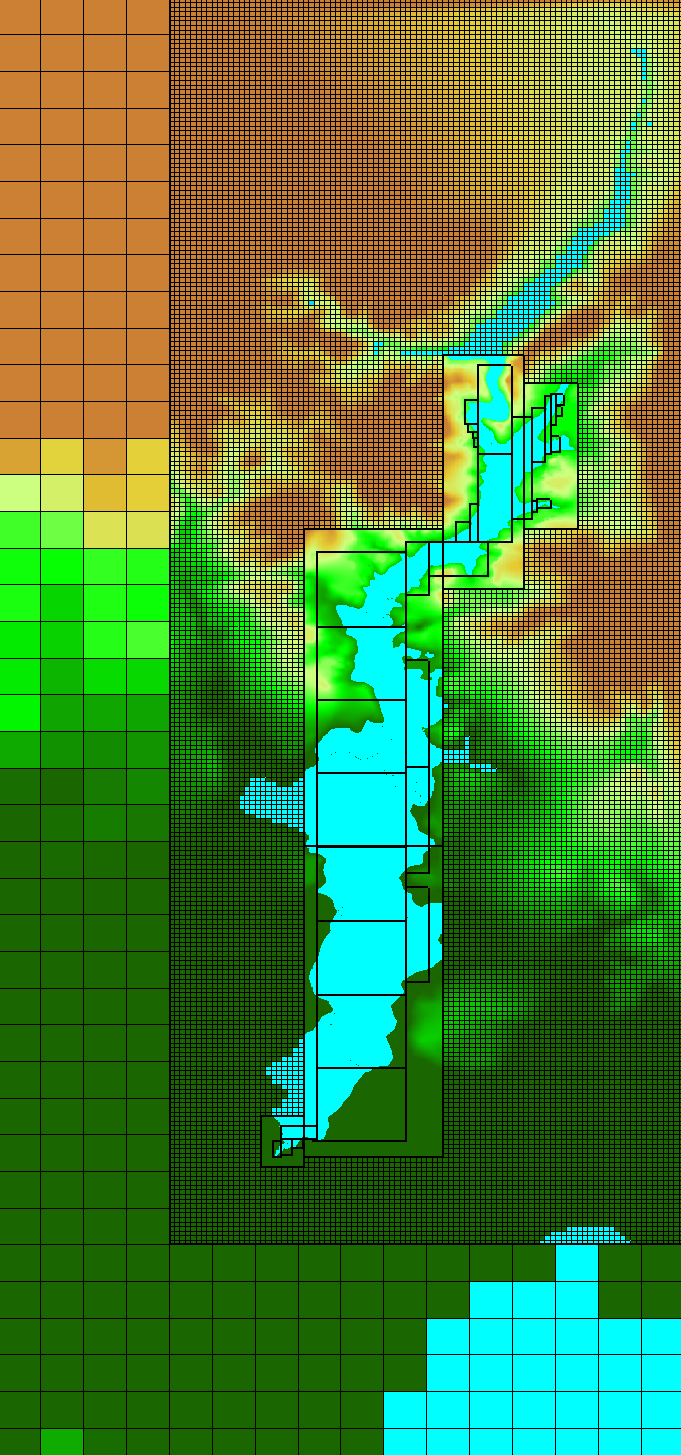}
\end{tabular}
\caption{\label{fig:Malpasset}A GeoClaw simulation of the Malpasset
dam-break flood from \cite{George2010}. Initial
and later times are shown from left to right. A very coarse level
1 grid is used where the flood has yet to arrive. The reservoir in the northeast
corner (upper right) is resolved on a level 2 grid. Level 3 and 4 grids surround
and evolve with the flood as it winds down a ravine, eventually entering the Reyran River Valley.
Individual level 3 and 4 grids are outlined; their grid lines are omitted
for clarity.}
\end{figure*}

\begin{figure*}
\begin{tabular}{c c}
\includegraphics[width=7.8cm]{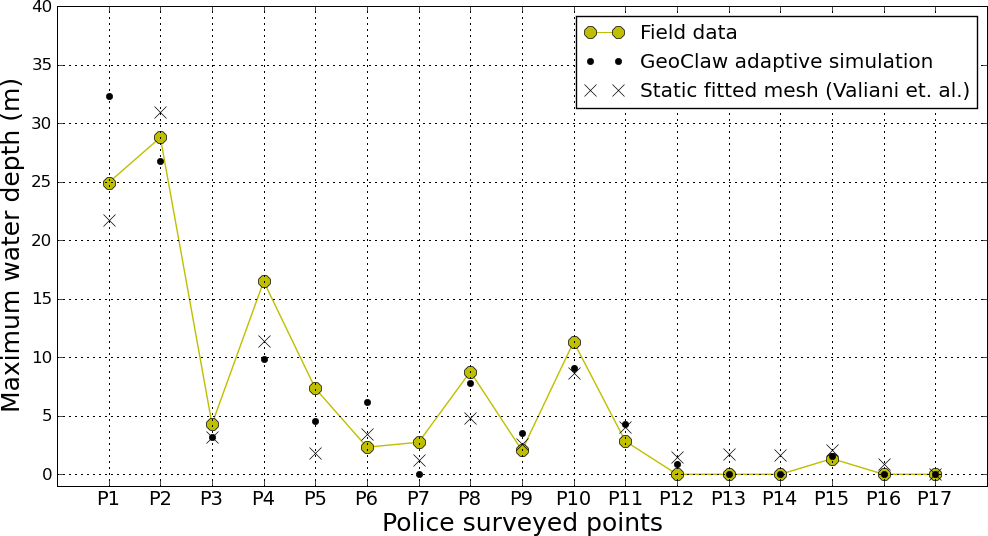} &
\includegraphics[width=7.8cm]{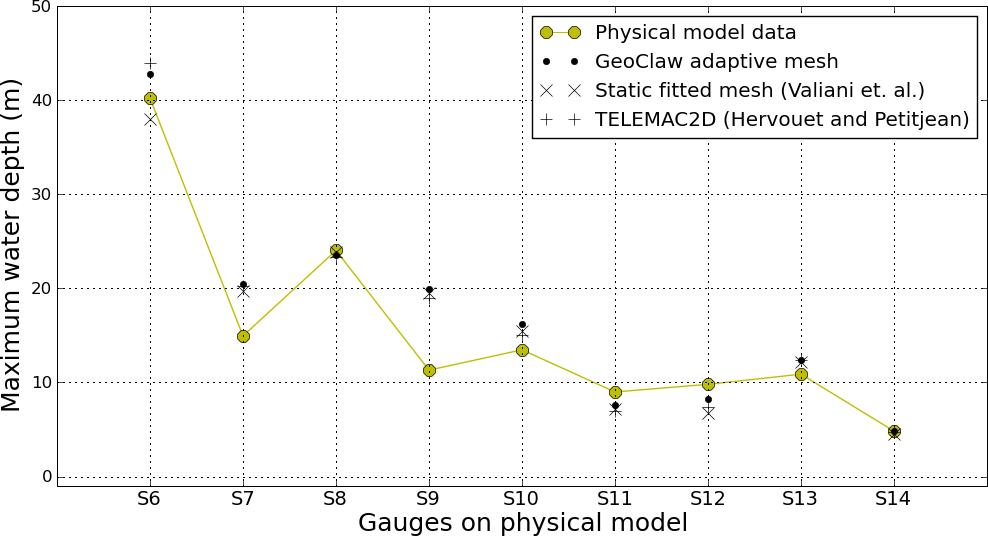}
\end{tabular}
\caption{\label{fig:Malpasset2} Comparison of the maximum water depths produced
by a GeoClaw simulation from \cite{George2010} with those presented by other authors
\cite{ValianiCaleffiEtAl2002,HervouetPetitjean1999}
using static topography-fit irregular meshes. Simulation results are compared
with field data for high-water marks collected after the flood at 17 surveyed
points (left), and with a physical scale-model experiment, where the maximum
water level was recorded by electronic gauges at 12 locations (right).
\protect\newstuff{Depths
not computed by GeoClaw are estimated by subtracting interpolated topography
from the reported surface elevations.  Figures
are adapted from \cite{George2010}, where total surface elevations are
shown.}
}
\end{figure*}

We have tested GeoClaw in this context by modeling the historic Malpasset
dam-break flood, which occurred in southern France in 1959 (see
\cite{George2010}). This thin-arch dam failed suddenly and explosively, sending
a roughly 60 m deep flood wave into the winding ravine below, eventually
inundating the Reyran River Valley. This disaster has served as a valuable test
case for code validation due to the extensive field data, such as high water
marks, collected after the event. For this problem we used a 6.47 km east--west
($x$-direction) by 16.58 km north--south ($y$-direction), rectangular grid. The
coarsest level 1 grid was 16 by 40 grid cells respectively ($\Delta x \times
\Delta y\approx 404.4 \mathrm{m} \times 414.5 \mathrm{m})$ . Level 2, level 3
and level 4 grids were then used to refine the flowing water, with refinement
ratios of 8, 4 and 4, yielding $\approx 3 \mathrm{m} \times 3 \mathrm{m}$ grid
cells on the finest level. Some snapshots from the simulation are shown in
\Fig{Malpasset}.
\newstuff{The maximum water depth computed by GeoClaw at various points}
was compared to other codes and empirical field and model data, shown in
\Fig{Malpasset2}. A detailed explanation of this test problem and comparison
can be found in \cite{George2010}.

\par
\section{Conclusions and future plans}\label{sec:conclusions}

The GeoClaw software project grew out of
the TsunamiClaw code developed by one
of the authors in his 2006 PhD thesis \cite{George2006}, which itself grew out
of Clawpack. It has undergone
several more years of development and testing, primarily on tsunami simulation.
The version recently released with Clawpack 4.5 (in July, 2010) is fairly robust
and stable, but will continue to be developed and improved in the future.
We are also incorporating OpenMP into the code to take advantage of
multi-core shared memory computers.

A number of on-going projects by the authors make use of this software.
We are currently developing depth-averaged models for two-phase flows consisting
of granular-fluid mixtures, applicable to debris-flow floods and volcanic lahars
or mudslides (e.g.
\cite{GeorgeIverson2011,Iverson1997,DenlingerIverson2001,DenlingerIverson2004}). These
flows often occur in rugged mountainous regions, and present many of the same
difficulties as overland flooding in terms of domain geometry.
The shallow water equations are also often used to model storm surge, e.g.
\cite{Resio:2008,westerink:2008}, and we are currently investigating the use
of both standard single layer shallow water equations and also a
multi-layer versions of the code as a possible improvement.
The multi-layer version may also be useful for modeling
tsunamis generated by
submarine landslides, as done for example in \cite{Fernandezetal:2008},
although the
multilayer shallow water equations introduce a number of new mathematical and
numerical challenges (see e.g.
\cite{AbgrallKarni2009,Audusse:multilayer,ca-ga:gibraltar}).
In the future the GeoClaw webpage and the application gallery will show
some results from these new application areas.
Animations and some GeoClaw code to accompany the simulations presented in this
paper are available at \cite{awr10-web}.

\vskip 10pt
{\bf Acknowledgments.}
This research was supported in part by
DOE grant DE-FG02-88ER25053, AFOSR grant FA9550-06-1-0203,
NSF Grant DMS-0914942, ONR Grant N00014-09-1-0649, and the Founders Term Professorship
in Applied Mathematics at the University of Washington.

\bibliographystyle{plain}

\end{document}